\documentclass[preprint,preprintnumbers,showpacs,aps,prd,amssymb]{revtex4-1}

\usepackage{graphicx}
\usepackage{bm}
\usepackage{amsmath}
\usepackage{xcolor}
\def\calH{{\cal H}}
\def\calL{{\cal L}}
\def\calO{{\cal O}}

\def\calU{{\cal U}}

\def\Bbar{{\bar B}}

\def\fbar{{\bar f}}
\def\hbar{{\bar h}}

\def\sbar{{\bar s}}

\def\SM{{\rm SM}}
\def\Br{{\rm Br}}
\def\eff{{\rm eff}}

\def\dU{{d_\calU}}
\def\LU{{\Lambda_\calU}}

\def\nn{\nonumber}

\def\Ks{K^{(*)}} 
\def\mhat{{\hat m}}
\def\shat{{\hat s}}
\def\uhat{{\hat u}}
\def\DamuU{{\Delta a_\mu^{\calU_V}}}
\def\DaeU{{\Delta a_e^{\calU_V}}}
\def\DalU{{\Delta a_\ell^{\calU_V}}}
\begin{document}
\title{$R(\Ks)$ with vector unparticles}
\author{Jong-Phil Lee}
\email{jongphil7@gmail.com}
\affiliation{Sang-Huh College,
Konkuk University, Seoul 05029, Korea}

\begin{abstract}
We analyze the unparticle effects on the lepton flavor universality violating ratios in $b\to s$ transition, $R(\Ks)$.
We concentrate on the vector unparticles which contribute to the relevant Wilson coefficients $C_{9,10}$.
New effects appear in the form of some powers of the momentum transfer squared, 
which is a different feature from other new physics models like leptoquarks or $Z'$.
Constraints from $B_s$-$\Bbar_s$ mixing, $B_s\to\mu^+\mu^-$, and the electron/muon anomalous magnetic moments are considered.
It is found that the vector unparticles successfully explain $R(\Ks)$ with moderate values of model parameters.
\end{abstract}

\maketitle
%%%%%%%%%%%%%%%%%%%%%%%%%%%%%%%%%%%%%%%%%%%%%%%%%%%%%%%%%%%%%%%%%%%%%%
\section{Introduction}
%%%%%%%%%%%%%%%%%%%%%%%%%%%%%%%%%%%%%%%%%%%%%%%%%%%%%%%%%%%%%%%%%%%%%%
%
The standard model (SM) of particle physics has been very successful since its advent,
but we have looked for new physics (NP) beyond the SM so far.
Recently there appeared some clues for NP in the flavor sector.
Among them is the ratio of the branching ratio of $B\to\Ks\ell\ell$ decays, $R(\Ks)$, defined by
\begin{equation}
R(\Ks)\equiv\frac{\Br(B\to\Ks\mu^+\mu^-)}{\Br(B\to\Ks e^+e^-)}~.
\end{equation}
Since the $b\to s$ transition involves a flavor-changing neutral current (FCNC) and does not occur at tree level in the SM, 
it is a good playground to probe NP.
Recently the LHCb Collaboration reported the updated measurements \cite{LHCb2103,LHCb1705,Geng2103},
\begin{eqnarray}
R(K)[1.1,6.0] &=& 0.846^{+0.044}_{-0.041}~,  \nn\\
R(K^*)[0.045,1.1] &=& 0.660^{+0.110}_{-0.070}\pm0.024~, \nn\\
R(K^*)[1.1,6.0] &=& 0.685^{+0.113}_{-0.069}\pm0.047~,
\label{RKs_EXP}
\end{eqnarray}
where the numbers in the square bracket are the bins of the momentum squared in ${\rm GeV}^2$.
According to the SM the ratios are very close to unity \cite{Hiller0310,Bobeth0709,Geng1704},
\begin{eqnarray}
R(K)_\SM[1.0,6.0] &=&1.0004^{+0.0008}_{-0.0007}~, \nn\\
R(K^*)_\SM[0.045,1.1] &=& 0.920^{+0.007}_{-0.006}~, \nn\\
R(K^*)_\SM[1.1,6.0] &=& 0.996^{+0.002}_{-0.002}~.
\label{RKs_SM}
\end{eqnarray}
The mismatch between the experiments and the SM indicates a compelling hint for the lepton-universality violation and NP.
There have been many proposals to solve the discrepancy including 
the leptoquark \cite{Hiller1408,Dorsner1603,Bauer1511,Chen1703,Crivellin1703,Calibbi1709,Blanke1801,Nomura2104,Angelescu2103,Du2104}, 
$Z'$ model \cite{Crivellin1501,Crivellin1503,Chiang1706,King1706,Chivukula1706,Cen2104,Davighi2105}, 
two-Higgs doublet model \cite{Hu1612,Crivellin1903,Rose1903}.
and so on \cite{Altmannshofer2002}.
\par
Among the various NP models the unparticle scenario is the most bizarre in that it involves the fractional number of ordinary particles
\cite{Georgi,unitarity}.
In this scenario it is assumed that there is a scale-invariant sector at very high energy which couples to the SM particles weakly
at some high scale $\LU$.
The effective field description of the scale-invariant sector at low energy regime is the unparticles.
%%%%%%%%%%%    %%%%%%%%%%%%%%%%%%%%%%%%%%%%%%%%%%%%%%%%%%%%
Assume that at some high energy $\sim M_\calU$ there is a scale-invariant ultraviolet (UV) theory in the hidden sector.
The formalism of the effective field theory is useful for describing the interaction between the UV theory and the SM sector.
For an SM operator $\calO_\SM$ and a UV operator $\calO_{\rm UV}$ the interaction below $M_\calU$ scale can be
written as $\calO_{\rm SM}\calO_{\rm UV}/M_\calU^{d_{\rm SM}+d_{\rm UV}-4}$,
where $d_{\rm UV(SM)}$ is the scaling dimension of $\calO_{\rm UV(SM)}$.
Through the renormalization flow one goes down along the scale until a new scale of $\Lambda_\calU$ appears 
through the dimensional transmutation where the scale invariance emerges.
Below $\Lambda_\calU$ one can match the theory with the new unparticle operator $\calO_\calU$ as \cite{Georgi}
\begin{equation}
C_\calU\frac{\Lambda_\calU^{d_{\rm UV}-\dU}}{M_\calU^{d_{SM}+d_{\rm UV}-4}}
\calO_{SM}\calO_{\calU}~,
\end{equation}
where $\dU$ is the scaling dimension of $\calO_\calU$ and $C_\calU$ is the matching coefficient.
The spectral function of the unparticle is given by the two-point function of $\calO_\calU$:
\begin{eqnarray}
\rho_\calU(P^2)&=&\int d^4x~e^{iP\cdot x}
\langle 0|\calO_\calU(x)\calO_\calU^\dagger(0)|0\rangle\nn\\
&=&
A_{\dU}\theta(P^0)\theta(P^2)(P^2)^{\dU-2}~,
\label{rhoU}
\end{eqnarray}
%%%%%%%%%%%%%%%%%%%%%%%%%%%%%%%%%%%%%%%%%%%%%%%%%%%%%%
where
\begin{equation}
A_{\dU}=\frac{16\pi^2\sqrt{\pi}}{(2\pi)^{2\dU}}
\frac{\Gamma(\dU+\frac{1}{2})}{\Gamma(\dU-1)\Gamma(2\dU)}~,
\end{equation}
is the normalization factor.
%%%%%%%%%%%%%%%%%%%%%%%%%%%%%%%%%%%%%%%%%%%%%%%%%%%%%%%%%%%%%%%
The scaling dimension $d_\calU$ needs not be integers in general by the scale invariance, 
and the unparticle seems like a $d_\calU$ number of massless particles \cite{Cheung0704}.
\par
After the first suggestion of unparticles many aspects of them are studied in
various collider phenomenology \cite{Cheung0704,Georgi2,Cheung0706,Kathrein1012}, 
dark matter \cite{Deshpande0707,Kikuchi0711,Gong0803,Jamil1107}, 
cosmology/astrophysics \cite{Davoudiasl0705,Freitas0708,Das0709}, 
and ungravity \cite{Mureika0712,Mureika0808,Gaete1005,Mureika1006,Mureika0909,JPL1106} to name a few.
Interestingly, unparticles are holographic dual of massless particles in higher dimensions \cite{Stephanov,JPL07}.
Recent applications include a remedy for the information loss in the black hole \cite{Rahaman},
emergent dark energy \cite{Artymowski}, $W$-pair production \cite{Soa}, and so on.
%%%%%%%%%%% %%%%%%%%%%%%%%%%%%%%%%%%%%%%%%%%%%%%%%%%%%%%%%%%%%%%%
In addition, there have been many studies of the unparticle effects on $B$ physics in various ways \cite{Geng0705,Mohanta08,He0805},
including $B_s$-${\bar B}_s$ mixing \cite{Lenz07,Mohanta07,Parry08,JPL1009} (for other mesons in \cite{Li07,Chen09}),
$B_s\to\mu^+\mu^-$ \cite{JPL1303}, $B\to D^{(*)}\tau\nu$ \cite{JPL1711,JPL2012}, etc.
\par
In principle there can be scalar and vector unparticles as well. 
But for the $R(\Ks)$ anomaly we focus on the vector unparticles 
because it is known that the relevant operators are only $\calO_9$ and $\calO_{10}$ as discussed in \cite{Alonso14,Geng17,Geng21}
\begin{eqnarray}
\calO_9 &=& \frac{e^2}{16\pi^2}\left(\sbar\gamma^\mu P_L b\right)\left({\bar\ell}\gamma_\mu\ell\right)~,\nn\\
\calO_{10} &=& \frac{e^2}{16\pi^2}\left(\sbar\gamma^\mu P_L b\right)\left({\bar\ell}\gamma_\mu\gamma_5\ell\right)~.
\label{O9O10}
\end{eqnarray}
The Lagrangian for the vector unparticle $\calO_\calU^\mu$ is
\begin{equation}
\calL_\calU^V 
= \frac{c_L^{V12}}{\LU^{d_V}}\fbar_2\gamma_\mu(1-\gamma_5)f_1\calO_\calU^\mu
+ \frac{c_R^{V12}}{\LU^{d_V}}\fbar_2\gamma_\mu(1+\gamma_5)f_1\calO_\calU^\mu~,
\end{equation}
where $c_{L,R}^{V12}$ are the vector couplings and $d_V$ is the scaling dimension of $\calO_\calU^\mu$.
In this analysis we assume that the vector unparticle mediates the FCNC at the tree level.
For flavor-conserving unparticles, see \cite{Bashiry08}.
We also restrict ourselves to only the left-handed couplings for simplicity.
\par
Typically unparticle contributions appear in the form of 
\begin{equation}
\sim {\cal A}\left(\frac{m_B^2}{\LU^2}\right)^{d_S} + {\cal B}\left(\frac{m_B^2}{\LU^2}\right)^{d_V-1}~,
\label{USUV}
\end{equation}
where $d_S$ is the scaling dimension of the scalar unparticle.
The unitarity condition \cite{unitarity} restricts the range of $d_{S,V}$ as
\begin{equation}
1\le d_S~,~~~3\le d_V~,
\label{unitarity}
\end{equation}
which suppresses the vector contribution of Eq.\ (\ref{USUV}) by a factor of $(m_B^2/\LU^2)$.
%%%%%%%%%%%   %%%%%%%%%%%%%%%%%%%%%%%%%%%%%%%%%%%%%%%%%%%%
\par
In this work we loose the condition of Eq.\ (\ref{unitarity}) and allow $1\le d_V\le 3$.
The propagator of the vector unparticle is proportional to $\sim -g^{\mu\nu}+[2(d_V-2)/(d_V-1)]P^\mu P^\nu/P^2$ where
$P$ is the momentum transfer \cite{unitarity}.
The main constraint on $d_V$ comes from the second term, but in our case it does not contribute to the relevant Wilson coefficients.
If we neglect the second term, the bound $1\le d_V\le 3$ is allowed just as for the scalar unparticles.
In cases where $\calO_\calU^\mu$ is transverse or non-gauge-invariant, the unitarity bound of \cite{unitarity} can be alleviated \cite{Bashiry08}.
\par
%
%
%
%%%%%%%%%%%%%%%%%%%%%%%%%%%%%%%%%%%%%%%%%%%%%%%%%%%%%%%
%\textcolor{red}{issue2}
For this range of $d_V$  contributions from the scalar unparticle is suppressed by a factor of 
$(m_B^2/\LU^2)$ when $d_S\sim d_V$, as one can see in Eq.\ (\ref{USUV}).  
Also the operators involved with the scalar unparticles are 
$\sbar(1\pm\gamma_5)b~{\bar\ell}(1-\gamma_5)\ell$,
which are irrelevant in our analysis on $R(\Ks)$.
\par
%%%%%%%%%%%%%%%%%%%%%%%%%%%%%%%%%%%%%%%%%%%%%%%%%%%%%%%
%
%
%
An interesting point of unparticle contribution to $R(\Ks)$ is that the relevant Wilson coefficients involve a factor of $s^{d_V-2}$
where $s$ is the momentum transfer squared.
For other NP scenarios like leptoquarks or $Z'$ models, the Wilson coefficients are $\sim 1/(G_Fm_{\rm NP}^2)$ \cite{Alda1805}.
Compared to this form unparticle contributions are quite delicate with respect to the model parameters,
which makes the phenomenology very interesting.
The unparticle parameters are constrained by other $B$ physics and lepton $(g-2)_{e,\mu}$
\cite{FNAL2104,Davier1010,Davier1706,Davier1908,Aoyama2006,Aoyama1205,Lin2112}.
The $B_s$-$\Bbar_s$ mixing and $a_{e,\mu}=(g-2)_{e,\mu}/2$ put bounds on the couplings to quarks and leptons, respectively,
and $B_s\to\mu^+\mu^-$ does on both couplings.
We exploit these constraints to explore allowed regions of the parameter space of unparticles.
\par
The paper is organized as follows.
In Sec.\ II we provide the formalism for $R(\Ks)$ with vector unparticles. 
Constraints from $B_s$-$\Bbar_s$ mixing, $\Br(B_s\to\mu^+\mu^-)$ and the anomalous lepton $(g-2)_{e,\mu}$ are also given.
Section III shows the allowed regions of the model parameters and discusses our results.
We conclude in Sec.\ IV.
%
%%%%%%%%%%%%%%%%%%%%%%%%%%%%%%%%%%%%%%%%%%%%%%%%%%%%%%%%%%%%%%%%%%%%%%
\section{$b\to s\ell\ell$ transition }
%%%%%%%%%%%%%%%%%%%%%%%%%%%%%%%%%%%%%%%%%%%%%%%%%%%%%%%%%%%%%%%%%%%%%%
%
The effective Hamiltonian for $b\to s\ell^+\ell^-$ is
\begin{equation}
\calH_{\rm eff} = -\frac{4G_F}{\sqrt{2}}V_{tb}V_{ts}^*\sum_{i=9}^{10} C_i(\mu)\calO_i(\mu)~,
\end{equation}
where the operators are defined in Eq.\ (\ref{O9O10}). 
The matrix elements of $\calH_{\rm eff}$ introduce several form factors. 
The differential decay rates for $B\to\Ks\ell^+\ell^-$ are given by \cite{Chang2010}
\begin{eqnarray}
\frac{d\Gamma_K}{d\shat}&=&
\frac{G_F^2\alpha^2m_B^5}{2^{10}\pi^5}|V_{tb}V_{ts}^*|^2\uhat_{K,\ell}\left\{
(|A'|^2+|C'|^2)\left(\lambda_K-\frac{\uhat_{K,\ell}^2}{3}\right)
+|C'|^24\mhat_\ell^2(2+2\mhat_K^2-\shat) \right .\nn\\
&&\left. +{\rm Re}(C'D'^*)8\mhat_\ell^2(1-\mhat_K^2)+|D'|^24\mhat_\ell^2\shat\right\}~,
\\
\frac{d\Gamma_{K^*}}{d\shat}&=&
\frac{G_F^2\alpha^2m_B^5}{2^{10}\pi^5}|V_{tb}V_{ts}^*|^2\uhat_{K^*,\ell}\left\{
\frac{|A|^2}{3}\shat\lambda_{K^*}\left(1+\frac{2\mhat_\ell^2}{\shat}\right)
+|E|^2\shat\frac{\uhat_{K^*,\ell}^2}{3}\right.\nn\\
&&
+\frac{|B|^2}{4\mhat_{K^*}^2}\left[\lambda_{K^*}-\frac{\uhat_{K^*,\ell}t^2}{3}+8\mhat^2_{K^*}(\shat+2\mhat_\ell^2)\right]
+\frac{|F|^2}{4\mhat_{K^*}^2}\left[\lambda_{K^*}-\frac{\uhat_{K^*,\ell}^2}{3}+8\mhat_{K^*}^2(\shat-4\mhat_\ell^2)\right]\nn\\
&&
+\frac{\lambda_{K^*}|C|^2}{4\mhat_{K^*}^2}\left(\lambda_{K^*}-\frac{\uhat_{K^*,\ell}^2}{3}\right)
+\frac{\lambda|_{K^*}|G|^2}{4\mhat_{K^*}^2}\left[\lambda_{K^*}-\frac{\uhat_{K^*,\ell}^2}{3}
	+4\mhat_\ell^2(2+2\mhat_{K^*}^2-\shat)\right]\nn\\
&&
-\frac{{\rm Re}(BC^*)}{2\mhat_{K^*}^2}\left(\lambda_{K^*}-\frac{\uhat_{K^*,\ell}^2}{3}\right)(1-\mhat_{K^*}^2-\shat)\nn\\
&&
-\frac{{\rm Re}(FG^*)}{2\mhat_{K^*}^2}\left[\left(\lambda_{K^*}-\frac{\uhat_{K^*,\ell}^2}{3}\right)(1-\mhat_{K^*}^2
	-\shat)-4\mhat_\ell^2\lambda_{K^*}\right]\nn\\
&&\left.
-\frac{2\mhat_\ell^2}{\mhat_{K^*}^2}\lambda_{K^*}\left[{\rm Re}(FH^*)-{\rm Re}(GH^*)(1-\mhat_{K^*}^2)\right]
+\frac{\mhat_\ell^2}{\mhat_{K^*}^2}\shat\lambda_{K^*}|H|^2\right\}~,
\end{eqnarray}
where the kinematic variables are
\begin{eqnarray}
\shat&=&\frac{s}{m_B^2}~,~~~\mhat_i=\frac{m_i}{m_B}~,\\
\lambda_H&=&1+\mhat_H^4+\shat^2-2\shat-2\mhat_H^2(1+\shat)~,~~~
\uhat_{H,\ell}=\sqrt{\lambda_H\left(1-\frac{4\mhat_\ell^2}{\shat}\right)}~,
\end{eqnarray}
and
\begin{equation}
s=(p_B-p_{\Ks})^2=(p_{\ell^+}+p_{\ell^-})^2~,
\end{equation}
is the momentum transfer squared.
Here the auxiliary functions $A',\cdots, D'$ and $A,\cdots, H$ involve the relevant Wilson coefficients and the form factors \cite{Ali99},
%
%%%%%%%%%%%    %%%%%%%%%%%%%%%%%%%%%%%%%%%%%%%%%%%%%%%%%%%
\begin{eqnarray}
A'&=&C_9 f_+ +\frac{2\mhat_b}{1+\mhat_K}C_7^\eff f_T ~,\\
B'&=&C_9 f_- -\frac{2\mhat_b}{\shat}(1-\mhat_K)C_7^\eff f_T ~,\\
C'&=&C_{10} f_+ ~,\\
D'&=&C_{10} f_- ~,
\end{eqnarray}
and
\begin{eqnarray}
A&=&\frac{2}{1+\mhat_{K^*}}C_9 V+\frac{4\mhat_b}{\shat}C_7^\eff T_1~,\\
B&=&(1+\mhat_{K^*})\left[C_9 A_1+\frac{2\mhat_b}{\shat}(1-\mhat_{K^*})C_7^\eff T_2\right]~,\\ 
C&=&\frac{1}{1-\mhat_{K^*}^2}\left[(1-\mhat_{K^*})C_9 A_2
	+2\mhat_b C_7^\eff \left(T_3+\frac{1-\mhat_{K^*}^2}{\shat}T_2\right)\right]~,\\
D&=&\frac{1}{\shat}\left\{C_9\left[(1+\mhat_{K^*})A_1-(1-\mhat_{K^*})A_2
	-2\mhat_{K^*}A_0\right]-2\mhat_b C_7^\eff T_3\right\}~,\\
E&=&\frac{2}{1+\mhat_{K^*}}C_{10} V ~,\\
F&=&(1+\mhat_{K^*})C_{10} A_1 ~,\\
G&=&\frac{1}{1+\mhat_{K^*}}C_{10} A_2 ~,\\
H&=&\frac{1}{\shat}C_{10}\left[(1+\mhat_{K^*})A_1-(1-\mhat_{K^*})A_2-2\mhat_{K^*} A_0\right] ~.
\end{eqnarray}
The form factors$f_{+,0,T}(s), ~A_{0,1,2}(s), ~T_{1,2,3}(s)$ and $V(s)$ are defined by ($q=p_B-p$)
\begin{eqnarray}
\langle K(p)|\sbar\gamma_\mu b|B(p_B)\rangle&=&
f_+\left[(p_B+p)_\mu-\frac{m_B^2-m_K^2}{s}q_\mu\right]+\frac{m_B^2-m_K^2}{s}f_0 q_\mu~,\\
\langle K(p)|\sbar\sigma_{\mu\nu} q^\nu(1+\gamma_5)b|B(p_B)\rangle&=&
i\left[(p_B+p)_\mu s -q_\mu(m_B^2-m_K^2)\right]\frac{f_T}{m_B+m_K}~,\\
\langle K^*(p)|(V-A)_\mu|B_(p_B)\rangle&=&
-i\epsilon_\mu^*(m_B+m_{K^*})A_1+i(p_B+p)_\mu(\epsilon^*\cdot p_B)\frac{A_2}{m_B+m_{K^*}}\nn\\
&&
+iq_\mu(\epsilon^*\cdot p_B)\frac{2m_{K^*}}{s}(A_3-A_0)
+\frac{\epsilon_{\mu\nu\rho\sigma}\epsilon^{*\nu}p_B^\rho b^\sigma}{m_B+m_{K^*}}2V~,
\end{eqnarray}
with $f_-=(f_0-f_+)(1-\mhat_K^2)/\shat$.
We adopt the exponential forms of \cite{Ali99} for the form factors. 
%
%
%
%%%%%%%%%%%%%%%%%%%%%%%%%%%%%%%%%%%%%%%%%%%
%\textcolor{red}{issue 5}
%%%%%%%%%%%
Here for $C_7^\eff$ we use the value of $C_7^{eff}=-0.313$ as in \cite{Ali99}.
%
%
%
%%%%%%%%%%%%%%%%%%%%%%%%%%%%%%%%%%%%%%%%%%%
%\par\textcolor{red}{issue 3}
%%%%%%%%%%%%%
\par
Contributions from the vector unparticles to the effective Hamiltonian are given by
\begin{eqnarray}
{\cal H}_{\rm eff}^{{\cal U}_V}/i
&=&
{\bar s}\left(\frac{ic_V^q}{\LU^{d_V-1}}\right)\gamma_\mu(1-\gamma_5)b
\left[\frac{iA_{d_V}e^{-i d_V}}{2\sin d_V\pi}\frac{1}{s^{2-d_V}}
     \left(-g^{\mu\nu}+\kappa_V\frac{q^\mu q^\nu}{s}\right)\right]  \nn\\
&&\times {\bar u}\left(\frac{ic_V^\ell}{\LU^{d_V-1}}\right)\gamma_\nu(1-\gamma_5) v~,     
\end{eqnarray}
where $\kappa_V=2(d_V-2)/(d_V-1)$.
As discussed in the Introduction we neglect the $\kappa_V$ term, and the result is
\begin{equation}
{\cal H}_{\rm eff}^{\calU_V}=
-\frac{A_{d_V}}{2\sin d_V\pi}\frac{e^{-i d_V\pi}}{s^{2-d_V}}
\frac{2c_V^q c_V^\ell}{\LU^{2 d_V-2}}
\frac{(4\pi)^2}{e^2}(\calO_9-\calO_{10})~.
\end{equation}
%%%%%%%%%%%%%%%%%%%%%%%%%%%%%%%%%%%%%%%%%%%
%
%
%
Now the vector unparticles contribute to the Wilson coefficients $C_{9,10}^{\calU_V}$ as follows:
\begin{equation}
C_9^{\calU_V} = -C_{10}^{\calU_V}
=\left(\frac{\sqrt{2}}{4G_F V_{tb}V_{ts}^*}\frac{16\pi^2}{e^2}\right)
\frac{A_{d_V}e^{-id_V\pi}}{\sin d_V\pi}(c_V^q c_V^\ell)\frac{s^{d_V-2}}{\Lambda_\calU^{2d_V-2}}~.
\end{equation}
We allow $c_V^{e,\mu}\ne 0$ so that NP effects could appear in both numerator and denominator of $R(\Ks)$.
\par
Constraints for the model parameters mainly come from $B_s\to\mu^+\mu^-$ decay and $B_s$-$\Bbar_s$ mixing.
The branching ratio of the $B_s\to\mu^+\mu^-$ with the vector unparticle is
\begin{equation}
\Br(B_s\to\mu^+\mu^-)_\calU 
= \Br(B_s\to\mu^+\mu^-)_\SM \left|1
	+\frac{C_{10}^{\calU_V}(B_s\to\mu^+\mu^-)}{C_{10}^\SM(B_s\to\mu^+\mu^-)}\right|^2~,
\end{equation}
where 
\begin{equation}
C_{10}^{\calU_V}(B_s\to\mu^+\mu^-) 
= \frac{\sqrt{2}\pi}{m_{B_s}^2 G_F\alpha V_{tb}V_{ts}^*} \frac{A_{d_V}e^{-id_V\pi}}{\sin d_V\pi}
   \left(\frac{m_{B_s}^2}{\LU^2}\right)^{d_V-1}\left(c_V^q c_V^\mu\right)~.
\label{C10Bs2mumu}
\end{equation}
A combination of $(c_V^q c_V^\mu)$ entering the branching ratio manifests  the feature of $B_s\to\mu^+\mu^-$ process.
For the scalar unparticle effects, see \cite{JPL1303}.
The experimental result is \cite{Geng2103}
\begin{equation}
\Br(B_s\to\mu^+\mu^-)_{\rm exp} = (2.842\pm0.333)\times 10^{-9}~.
\end{equation}
\par
The $b\to s$ transition is also affected by the $B_s$-$\Bbar_s$ mixing.
The mass difference \cite{HFLAV1909} is measured to be
\begin{equation}
\Delta M_s^{\rm exp} = (17.757\pm0.021)/{\rm ps}~,
\label{DMs_exp}
\end{equation}
which would be compared to the SM prediction \cite{Luzio1909}
\begin{equation}
\Delta M_s^\SM = \left(18.4^{+0.7}_{-1.2}\right)/{\rm ps}~.
\label{DMs_SM}
\end{equation}
Unparticle effects on $\Delta M_s$ can be found in \cite{JPL1009}.
We only consider the vector unparticle contributions by switching off the scalar couplings.
\par
Vector unparticles can also contribute to the magnetic moment of leptons.
The amount is calculated to be \cite{Liao}
\begin{equation}
\Delta a_\ell^{\calU_V} 
= \frac{A_{d_V}}{8\pi^2\sin d_V\pi}
\frac{\Gamma(3-d_V)}{\Gamma(d_V+2)}\left[4\Gamma(2d_V-2)-\Gamma(2d_V-1)\right]
\left(c_V^\ell\right)^2 \left(\frac{m_\ell^2}{\LU^2}\right)^{d_V-1}~,
\label{aellV}
\end{equation}
for leptons $\ell =e, \mu$.
Recently the first results of the Fermilab National Accelerator Laboratory on $a_\mu$ is announced.
The new experimental average is \cite{FNAL2104}
\begin{equation}
a_\mu^{\rm exp} = 116\ 592\ 061(41)\times 10^{-11}~,
\end{equation}
which deviates from the SM value by $4.2\sigma$ \cite{Aoyama2006,Du2104,Davier1010,Davier1706,Davier1908}
\begin{equation}
\Delta a_\mu = a_\mu^{\rm exp}-a_\mu^\SM = (251\pm59)\times 10^{-11}.
\label{Damu}
\end{equation}
For the electron magnetic moment, we use the following result \cite{Aoyama1205,Lin2112}
\begin{equation}
\Delta a_e = a_e^{\rm exp}-a_e^\SM = (-106\pm82)\times 10^{-14}.
\label{Dae}
\end{equation}
%
%%%%%%%%%%%%%%%%%%%%%%%%%%%%%%%%%%%%%%%%%%%%%%%%%%%%%%%
\section{Results and Discussions}
%%%%%%%%%%%%%%%%%%%%%%%%%%%%%%%%%%%%%%%%%%%%%%%%%%%%%%%
%
As mentioned in the Introduction, we loose the unitarity constraint for $d_V$ and scan for $1\le d_V\le 3$ \cite{Bashiry08}.
%%%%%%%%%%%%%%%%%%%%%%%%%%%%%%%%%%%%%%%%%%%%%%
%
%
%
%\textcolor{red}{issue 1}
Constraints for the relevant parameters come from $B_s\to\mu^+\mu^-$ decay, 
$B_s$-${\bar B}_s$ mixing, and $b\to s\gamma$ decays.
But many works have been done for $d_V\ge 3$ where the unitarity constraint holds,
so the previous constraints for $d_V$ and $c_V^{q,\ell}$ would not affect current analysis strongly.
As mentioned in the Introduction for $d_V\ge 3$ contributions from the vector unparticles are 
quite suppressed, so large values of $c_V^{q,\ell}$ are allowed.
More specifically, the suppression factor is about $m_B^2/\LU^2\sim 3\times 10^{-5}$ for $\LU=1$ TeV,
therefore $c_V^q c_V^\ell\sim 10^5$ would be needed to catch up with the scalar unparticles.
In \cite{JPL1009} we analyzed the effects of both scalar and vector unparticles on $B_s$-${\bar B}_s$ mixing.
It was shown that the suppression factor for vector unparticles is effectively 
\begin{equation}
\sim \frac{1}{(2\pi)^4}\left(\frac{m_{B_s}^2}{\LU^2}\right)=1 .8\times10^{-8}~.
\end{equation}
for $\LU=1$ TeV, and the scalar unparticle is dominant.
In \cite{JPL1009} the mass difference as well as the mixing parameter are considered
(in this analysis we only consider the mass difference for simplicity),
and the relevant coupling of the scalar unparticle $c_S$ is in the range $0.1\lesssim |c_S|\lesssim 1$.
If we turn off $c_S$ and turn on only the vector unparticles and allow $d_V\ge 1$,
then we could get similar results for $c_V^q$.
As will be seen later, current analysis would provide a stronger bound on $c_V^q$.
%\textcolor{red}{issue 1-END}
%
%
%
\par
%%%%%%%%%%%%%%%%%%%%%%%%%%%%%%%%%%%%%%%%%%%%%%%
For higher values of $d_V$ the effects of $\calO_\calU^\mu$ are very suppressed.
%
%----------------- Figure 1 ------------------------------------------------
\begin{figure}
\begin{tabular}{cc}
\hspace{-3cm}\includegraphics[scale=0.17]{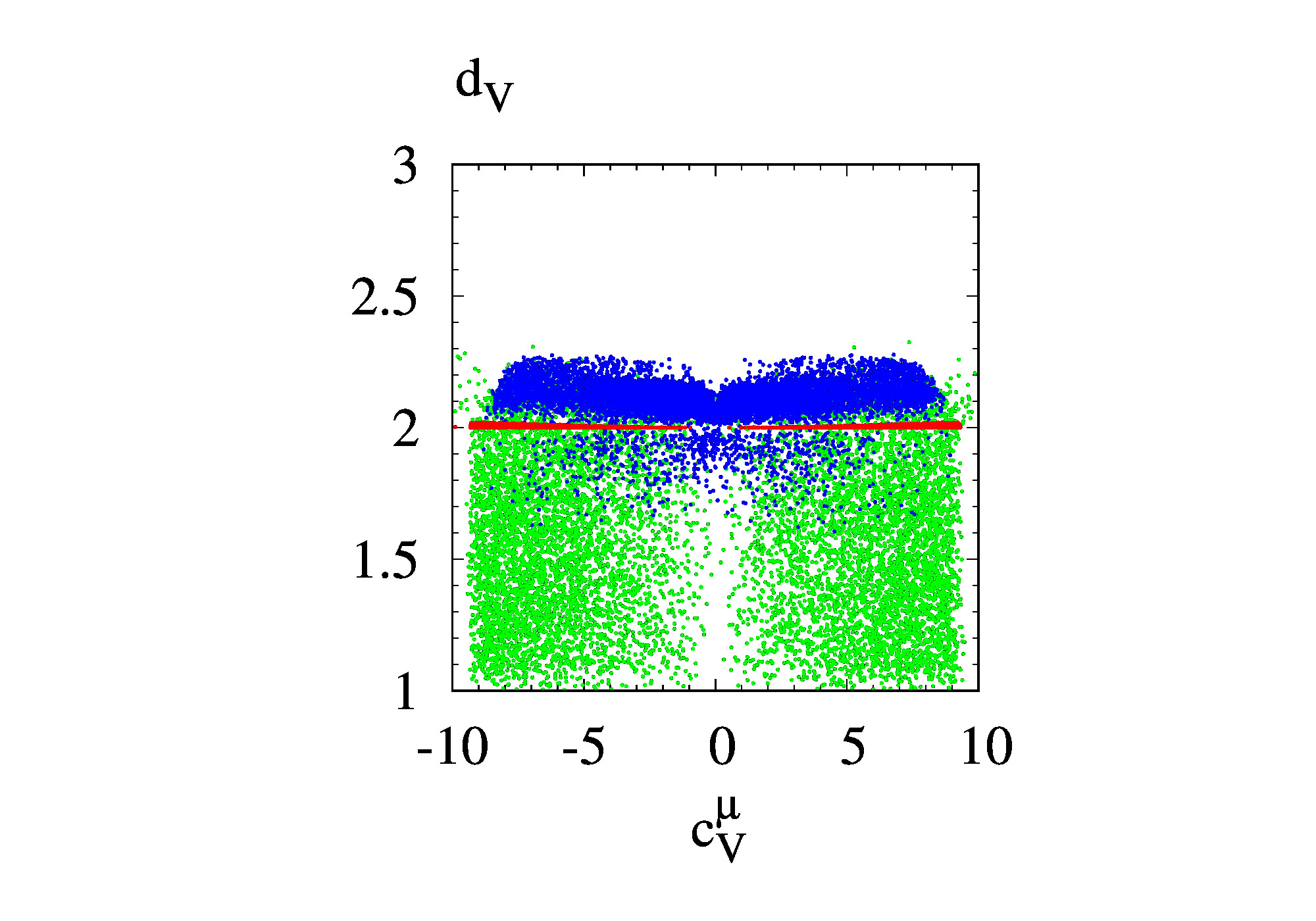} &
\hspace{-2cm}\includegraphics[scale=0.17]{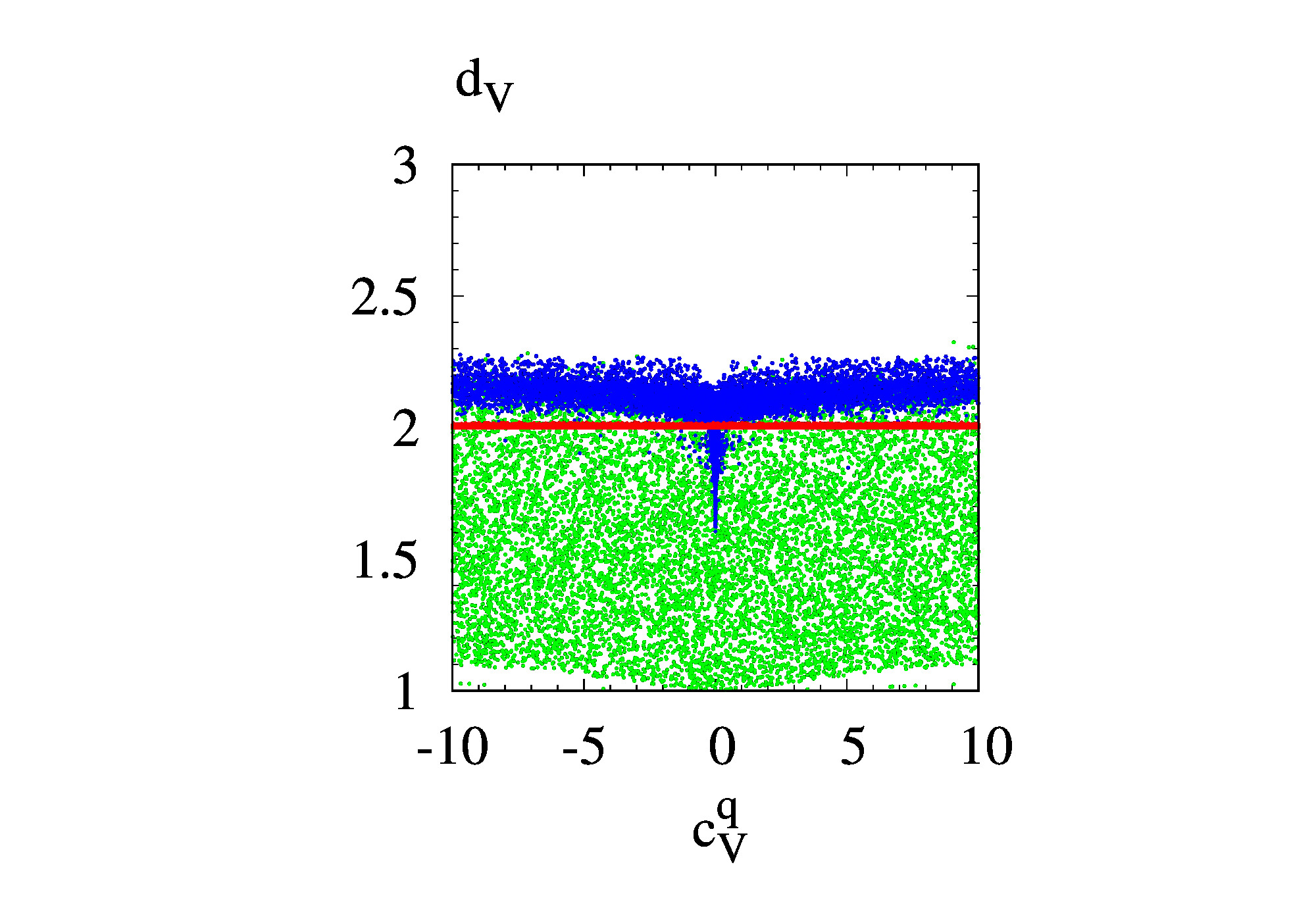} \\
\hspace{-3cm} (a) & \hspace{-2cm} (b) \\
\hspace{-3cm}\includegraphics[scale=0.17]{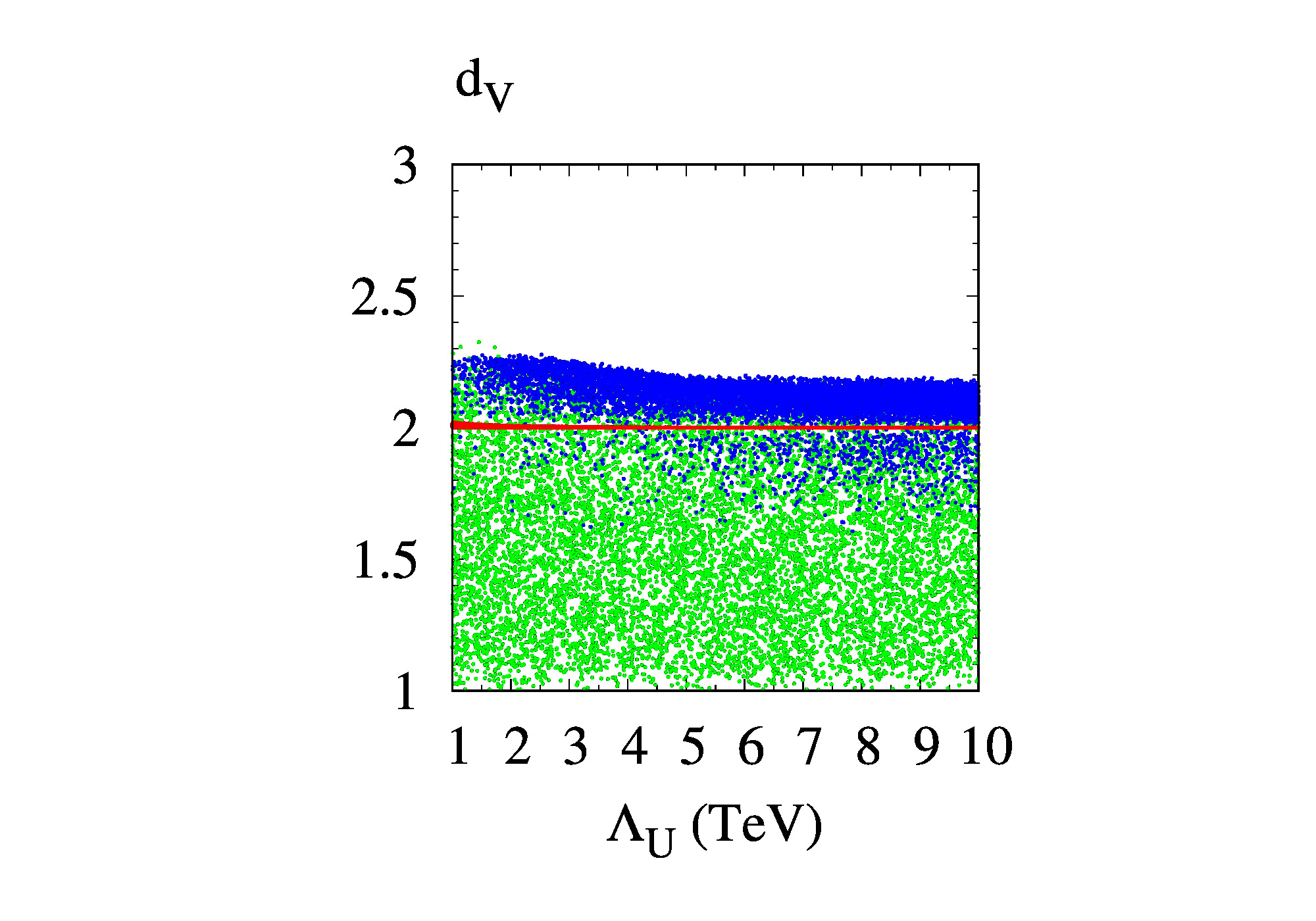} & \\
\hspace{-3cm} (c) & 
\end{tabular}
\caption{\label{F_dV} Allowed regions of model parameters 
(a) $d_V$ vs $c_V^\mu$, (b) $d_V$ vs $c_V^q$, and (c) $d_V$ vs $\LU$ (in TeV)
with the conditions 
$B_s$-$\Bbar_s$ mixing $+R(\Ks)$ (green), $\Delta a_{\ell}+R(\Ks)$ (red), $B_s\to\mu^+\mu^-+R(\Ks)$ (blue)
at the $2\sigma$ level.}
\end{figure}
%----------------------------------------------------------------------------
%
In Fig.\ \ref{F_dV} we show the allowed regions of $d_V$ vs $c_V^\mu$, $c_V^q$, and $\LU$ at the $2\sigma$ level.  
Constraints from the $B_s$-$\Bbar_s$ mixing are rather mild because 
the SM value of Eq.\ (\ref{DMs_SM}) is close to the experimental measurement of Eq.\ (\ref{DMs_exp}) 
and the uncertainties in the SM are still large .
The result is that green regions of figures in Fig.\ \ref{F_dV} are quite broad.
For larger values of SM, for example $\Delta M_s^\SM =(20.01\pm1.25) /{\rm ps}$ as in \cite{Luzio1712},
it would require the quark coupling $c_V^q$ to be larger which could make a tension with other constraints.
\par
%%%%%%%%%%%%%%%%%%%%%%%%%%%%%%%%%%%%%%%%%%%%%%%%%%%%%%%
%
%  			Updates
%
Other observables, $\Br(B_s\to\mu^+\mu^-)$ and $\Delta a_{\ell}$ provide with strong constraints on relevant parameters.
Especially $\Delta a_\ell$ puts stringent bound on $d_V$.
The reason is like this. 
In Eq.\ (\ref{aellV}), $\sin d_V\pi$ factor must be kept positive to meet the experimental data, $\Delta a_{e,\mu}$,
which allows only $2\le d_V \le 3$ .
But for larger values of $d_V$ the suppression from $(m_{\ell}^2/\LU^2)^{d_V -1}$ gets severe and cannot satisfy
the sizable $\Delta a_\mu$.
Thus $d_V$ stays in a narrow region around $\approx 2$.
It is straightforward that $\Delta a_{e(\mu)}$ affects $c_V^{e(\mu)}$ but not $c_V^q$ as shown in Figs.\ \ref{F_dV} (a) and (b).
Small values of $|c_V^{e(\mu)}|\ll 1$ are not allowed for $\Delta a_{e(\mu)}+R(\Ks)$ (red dots),
since they cannot make sizable, especially, $\Delta a_\mu$.
Similar pattern is expected for $c_V^e$.
On the other hand the narrow red band of Fig.\ \ref{F_dV} (b) is quite flat because $c_V^q$ is irrelevant to $\Delta a_{\ell}$.
\par
As for $B_s\to\mu^+\mu^-$ decay, the combined factor $c_V^q c_V^\mu$ contributes to the branching ratio.
But the allowed regions of $c_V^q$ and $c_V^\mu$ are different from each other.
The reason is that $c_V^q$ contributes to $R(\Ks)$ both in the numerator and the denominator 
while $c_V^\mu$ does only in the numerator.
Different patterns of $c_V^\mu$ from those of $c_V^q$ with respect to $d_V$ (blue dots in Fig.\ \ref{F_dV} (a) and (b))
come from the fact that in the ratios of $R(\Ks)$ effects of $c_V^\mu$ can be balanced by $c_V^e$.
For those values of $|c_V^q|\gtrsim 2$ only $d_V\gtrsim 2$ are allowed, while this is not the case for $c_V^\mu$.
Figure \ref{F_dV} (c) depicts the behavior of $d_V$ versus $\LU$.
As is common in the unparticle scenario, larger $d_V$ is allowed for smaller values of $\LU$.
For example, the factor of $(m_{B_s}^2/\LU^2)^{d_V}$ in Eq.\ (\ref{C10Bs2mumu}) for $B_s\to\mu^+\mu^-$ 
is very typical to result in the blue distributions in Fig.\ \ref{F_dV} (c).
For red dots from $\Delta a_\ell$ the allowed $d_V$ is very narrow near 2,
which does not provoke so strong suppression of $(m_{B_s}^2/\LU^2)^{d_V}$,
thus larger $\LU$ is also possible.

%For large values of both $d_V$ and $\LU$, relevant coupling $(c_V^q c_V^\mu)$ could compensate the suppression.
%
%----------------- Figure 2 ------------------------------------------------
\begin{figure}
\begin{tabular}{cc}
\hspace{-3cm}\includegraphics[scale=0.17]{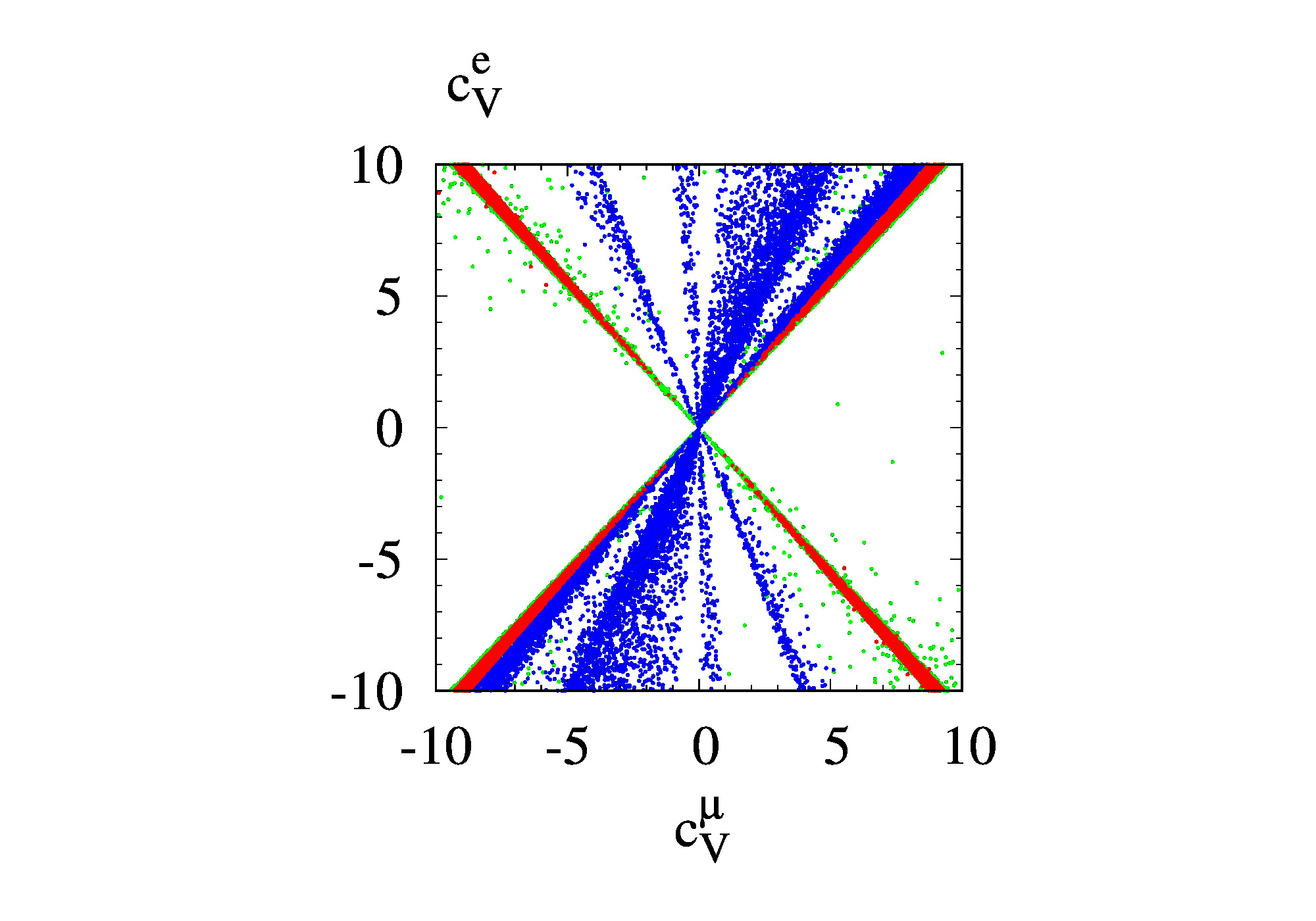} &
\hspace{-3cm}\includegraphics[scale=0.17]{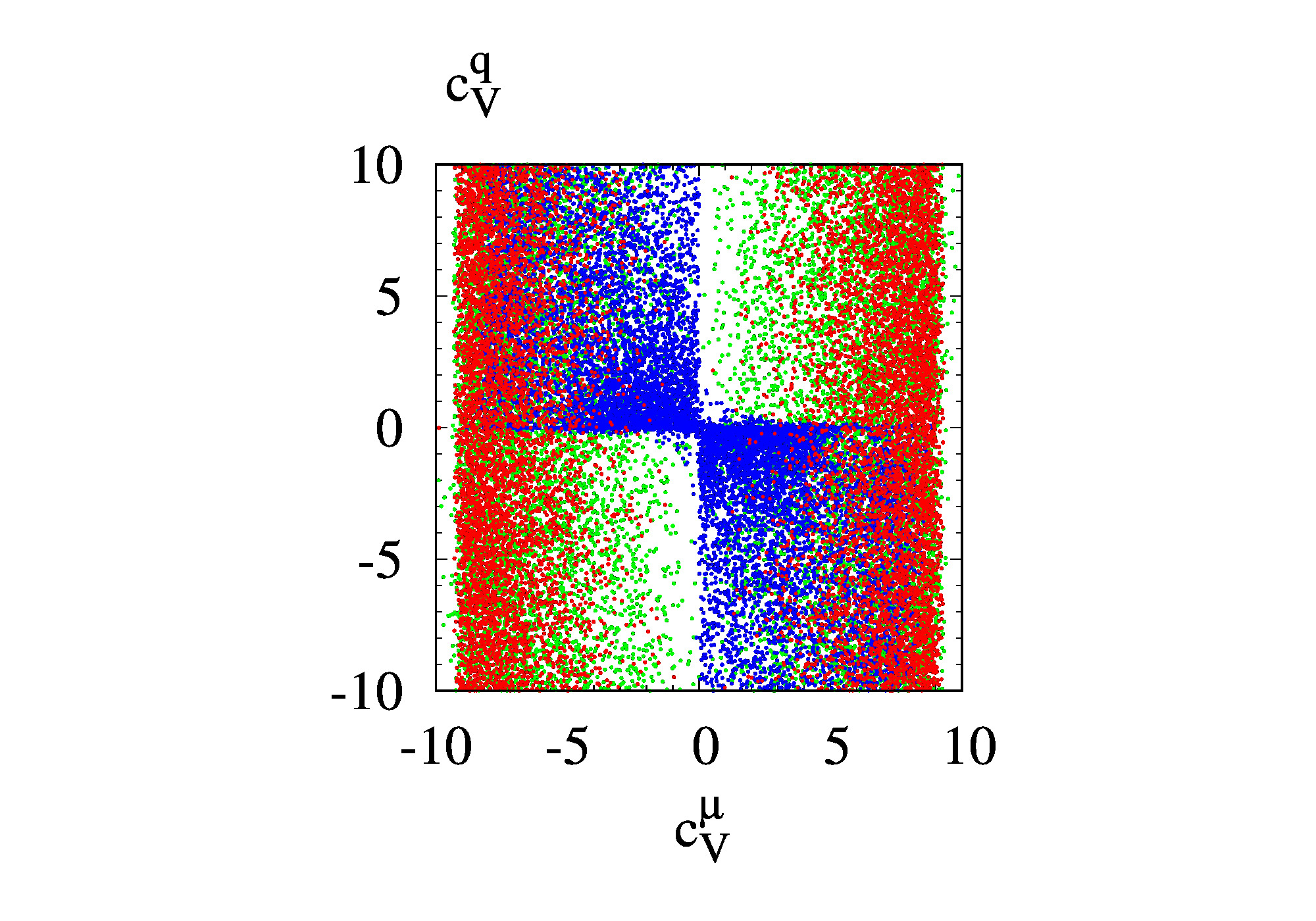} \\
\hspace{-3cm} (a)  & \hspace{-3cm} (b) 
\end{tabular}
\caption{\label{F_cV} Allowed regions of model parameters 
(a) $c_V^e$ vs $c_V^\mu$ and (b) $c_V^q$ vs $c_V^\mu$
with the conditions 
$B_s$-$\Bbar_s$ mixing $+R(\Ks)$ (green), $a_{e,\mu}+R(\Ks)$ (red), $B_s\to\mu^+\mu^-+R(\Ks)$ (blue)
at the $2\sigma$ level.
}
\end{figure}
%----------------------------------------------------------------------------
%
\par
Figure \ref{F_cV} shows the allowed regions for the couplings $c_V^q$, $c_V^\mu$, and $c_V^e$.
For $c_V^\mu$ and $c_V^e$ in Fig.\ \ref{F_cV} (a), constraints from $\Delta a_{e,\mu}$ (red dots) 
combined with $R(\Ks)$ are very strong.
As discussed above $\Delta a_{e,\mu}$ does not allow small values of $|c_V^{e,\mu}|\ll 1$.
Figure \ref{F_cV} (a) shows clearly that it is $\Delta a_{e,\mu}$ that prevents the region $|c_V^{e,\mu}|\ll 1$.
Since $c_V^e$ has nothing to do with $B_s$-$\Bbar_s$ mixing (green) nor $B_s\to\mu^+\mu^-$ (blue), 
one can infer that the "X"-like pattern of $c_V^e$ vs $c_V^\mu$ comes from $R(\Ks)$.
It means that the effects of $c_V^\mu$ on $R(\Ks)$ are balanced by $c_V^e$ to fit the measurements.
In view of $\DalU$, $\DaeU/\DamuU\sim (c_V^e/c_V^\mu)^2(m_e^2/m_\mu^2)^{d_V-1}\sim \calO(10^{-3})$.
It means that for $d_V\approx 2$, $c_V^e$ and $c_V^\mu$ must be nearly of the same order.
For larger values of $d_V$ one needs a big hierarchy between $c_V^e$ and $c_V^\mu$, 
which would spoil the $R(\Ks)$ fitting.
Figure \ref{F_cV} (b) shows the distribution of $c_V^q$ vs $c_V^\mu$.
We find that the feature of $(c_V^q c_V^\mu)$ for $\Br(B_s\to\mu^+\mu^-)$ (blue dots) is not so clear in this parameter space.
It means other parameters $d_V$ and $\LU$ would have more room to fit $\Br(B_s\to\mu^+\mu^-)$.
The quark coupling $c_V^q$ is not affected by $\Delta a_{e,\mu}$ (red dots) as expected.
Obviously $|c_V^\mu|\ll 1$ is disfavored in the red distribution.
\par
%----------------- Figure 3 ------------------------------------------------
\begin{figure}
\begin{tabular}{cc}
\hspace{-2cm}\includegraphics[scale=0.15]{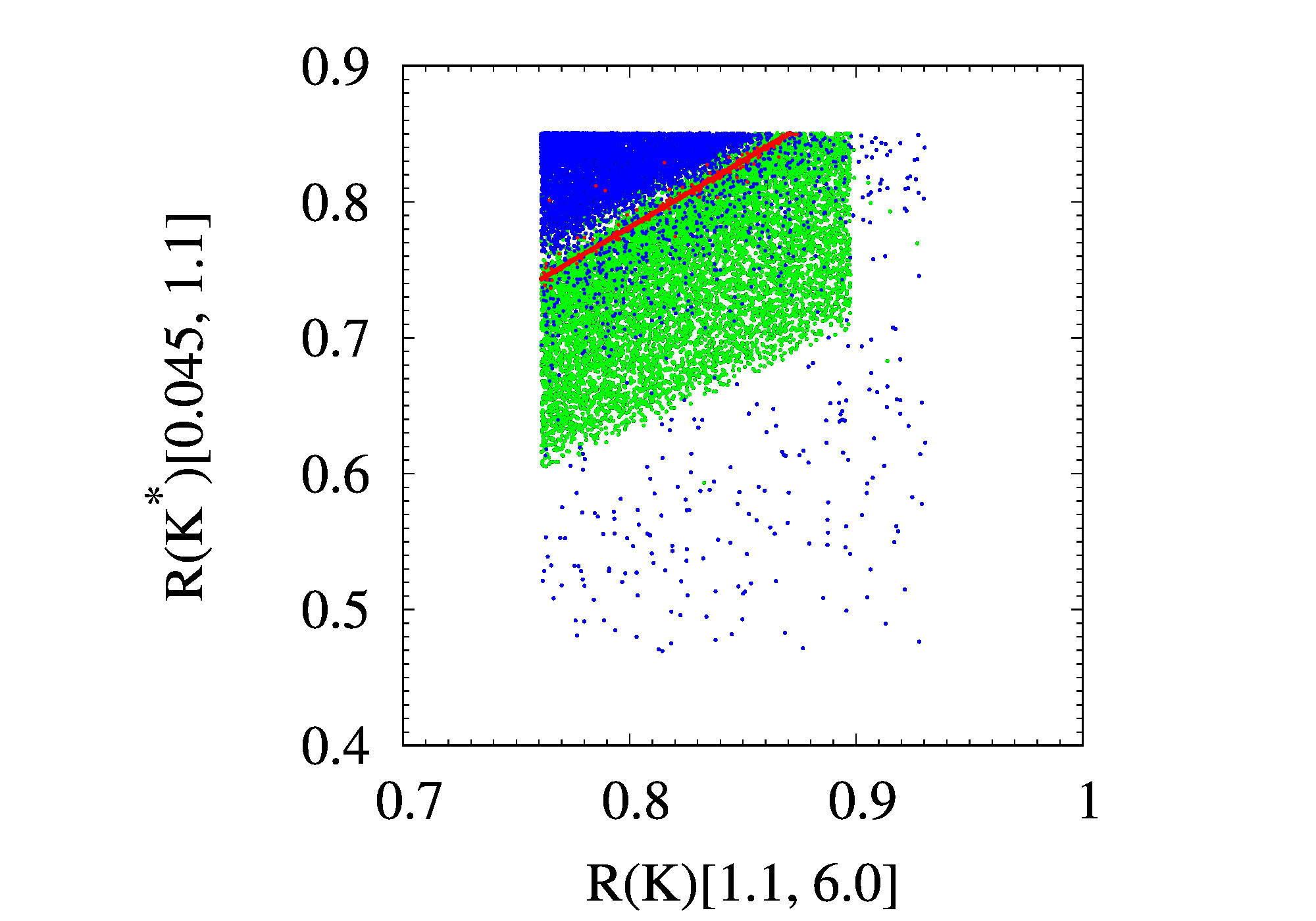}&
\hspace{-1cm}\includegraphics[scale=0.15]{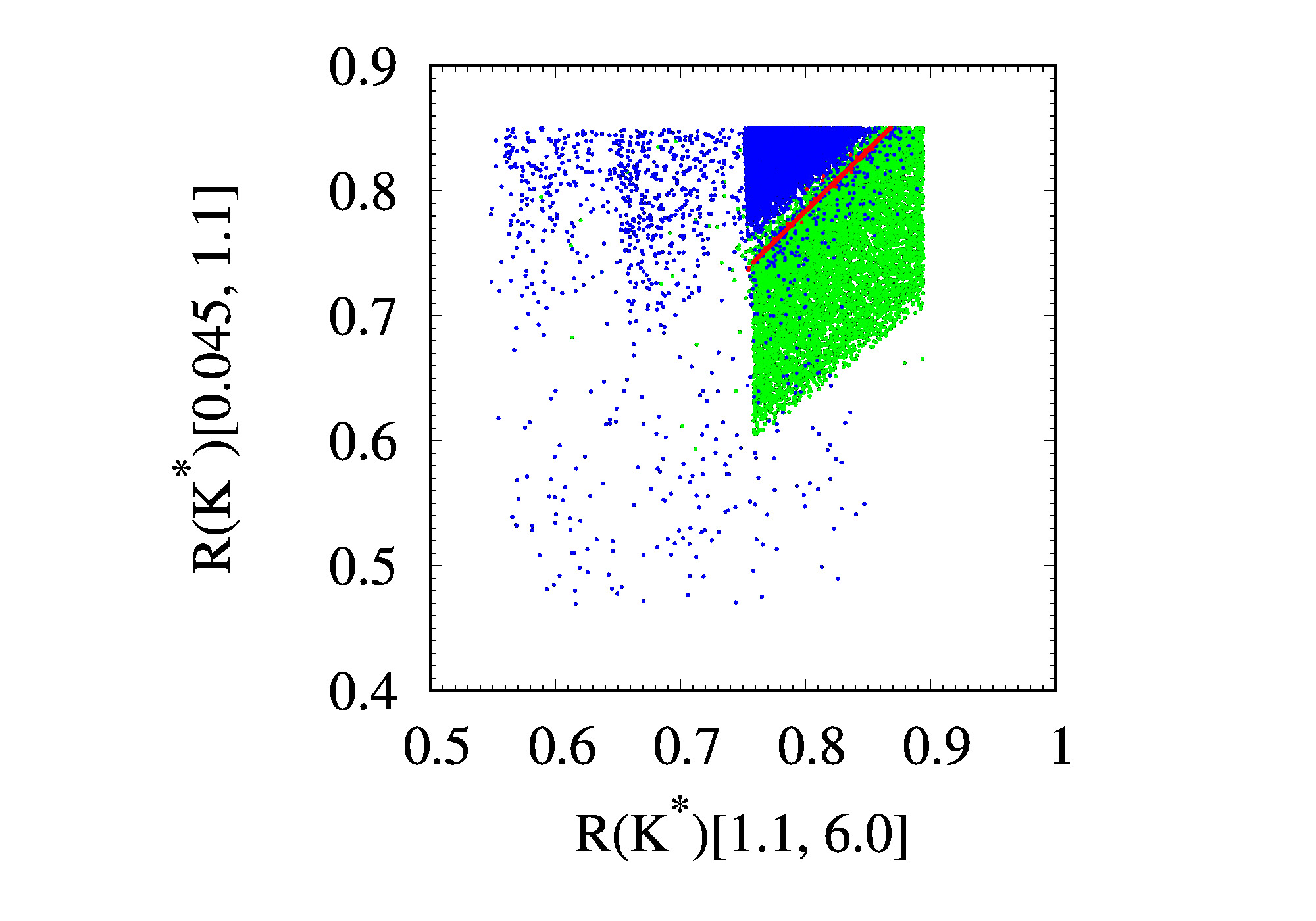}\\
\hspace{-2cm} (a) & \hspace{-1cm} (b) 
\end{tabular}
\caption{\label{F_RK} Allowed regions of 
(a) $R(K^*)[0.045:1.1]$ vs $R(K)[1.1:6.0]$ and
(b) $R(K^*)[0.045:1.1]$ vs $R(K^*)[1.1:6.0]$
with the constraints
$B_s$-$\Bbar_s$ mixing $+R(\Ks)$ (green), $a_{e,\mu}+R(\Ks)$ (red), $B_s\to\mu^+\mu^-+R(\Ks)$ (blue)
at the $2\sigma$ level.
}
\end{figure}
%----------------------------------------------------------------------------
%
In Fig.\ \ref{F_RK} we plot the allowed $R(\Ks)$ regions with different constraints.
Note that $\Delta a_{e,\mu}$ (red dots) puts the strongest bounds in $\calO_\calU^\mu$ scenario for $R(\Ks)$.
As discussed before,  $\Delta a_{e,\mu}$ provides a significant limit on $d_V$ and restricts directly the coupling $c_V^{e,\mu}$, 
while $B_s\to\mu^+\mu^-$ dose the combination $c_V^q c_V^\mu$, which allows more room for each coupling.
%
%----------------- Figure 4 ------------------------------------------------
\begin{figure}
\begin{tabular}{cc}
\hspace{-2cm}\includegraphics[scale=0.17]{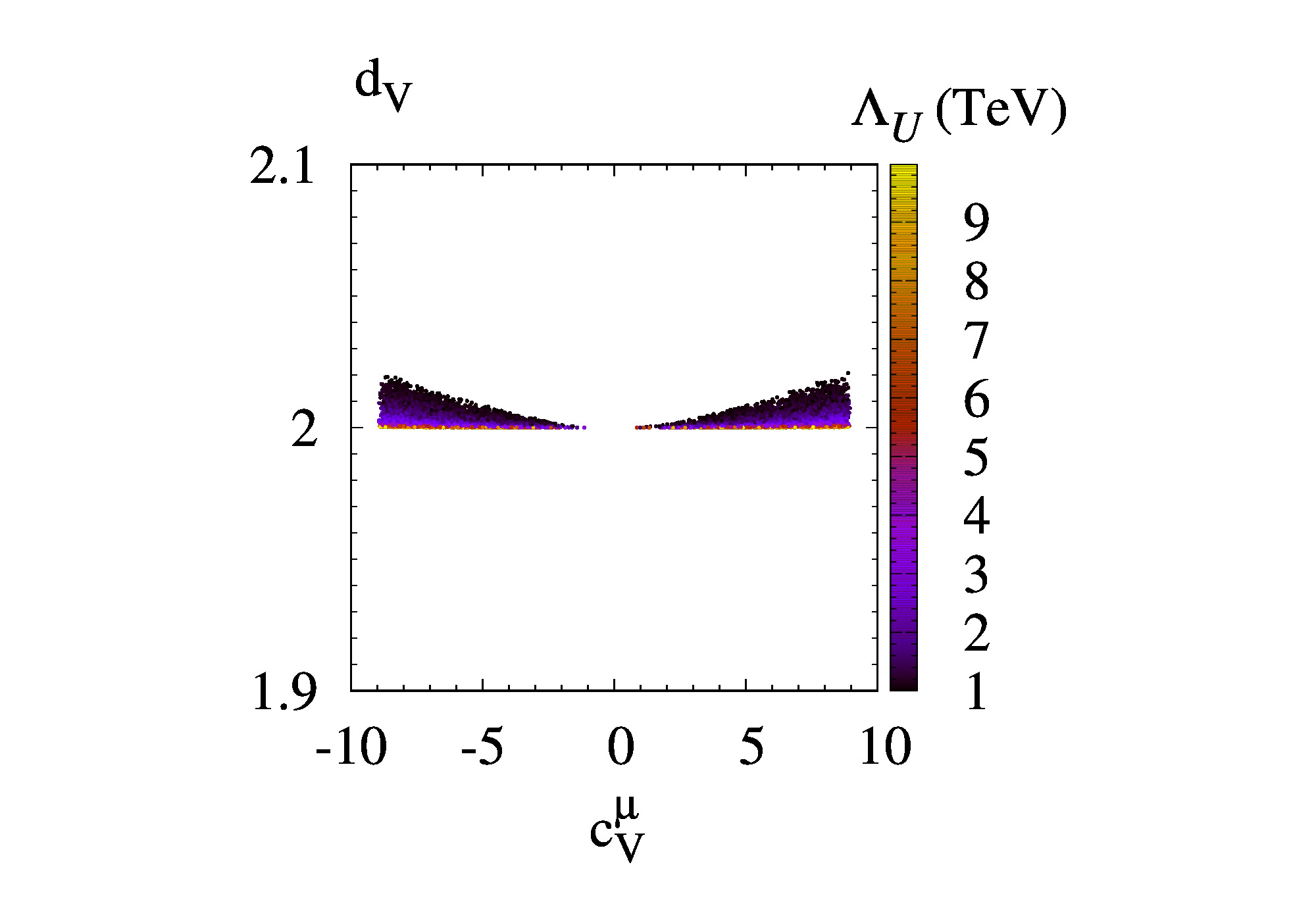}&
\hspace{-2cm}\includegraphics[scale=0.17]{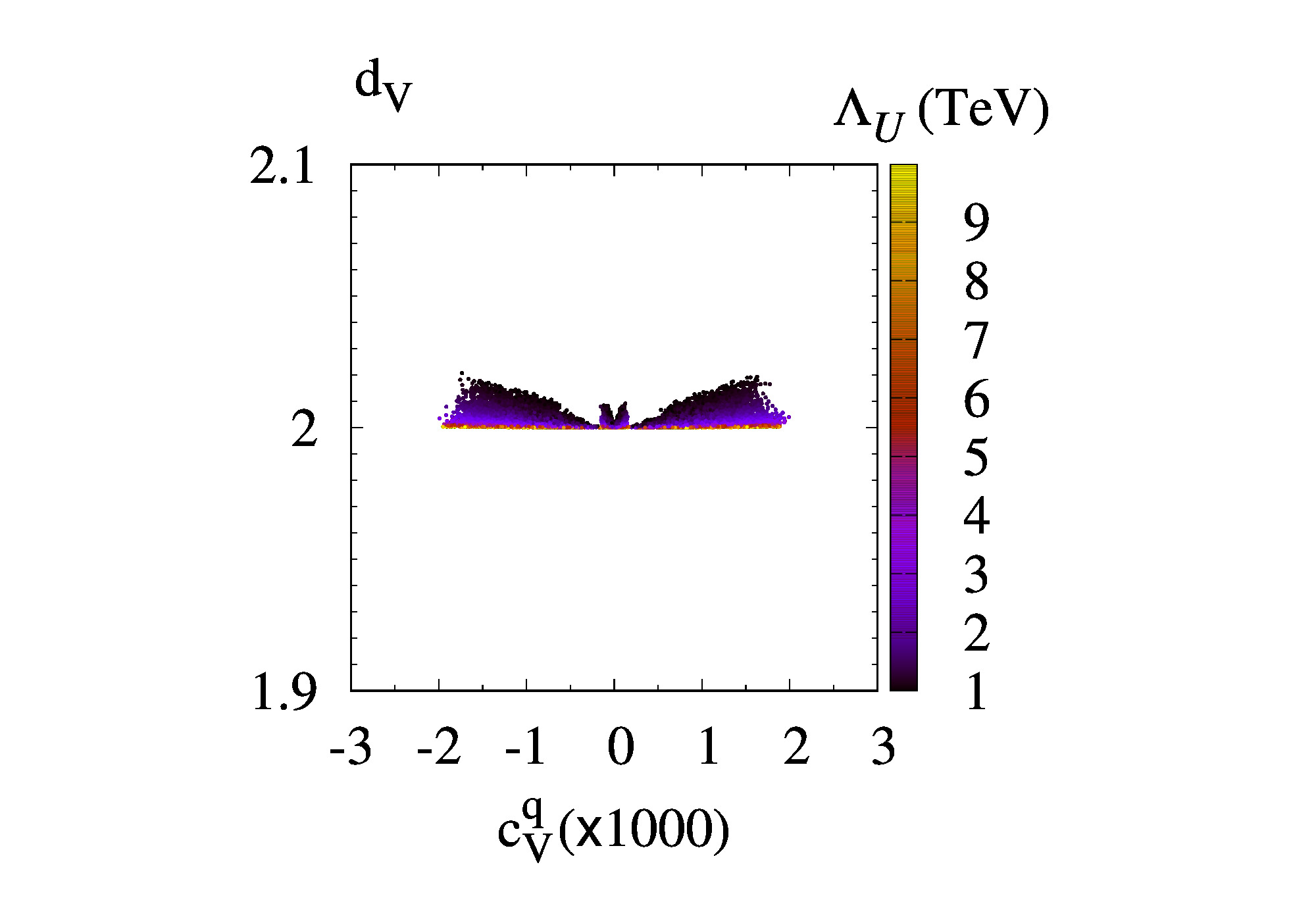}\\
\hspace{-2cm} (a) & \hspace{-2.cm} (b) 
\end{tabular}
\caption{\label{F_3d_dV} Allowed regions of (a) $d_V$ vs $c_V^\mu$ and (b) $d_V$ vs $c_V^q$ with respect to $\LU$,
with all the constraints at the $2\sigma$ level.
}
\end{figure}
%----------------------------------------------------------------------------
\par
\begin{figure}
\begin{tabular}{cc}
\hspace{-2cm}\includegraphics[scale=0.17]{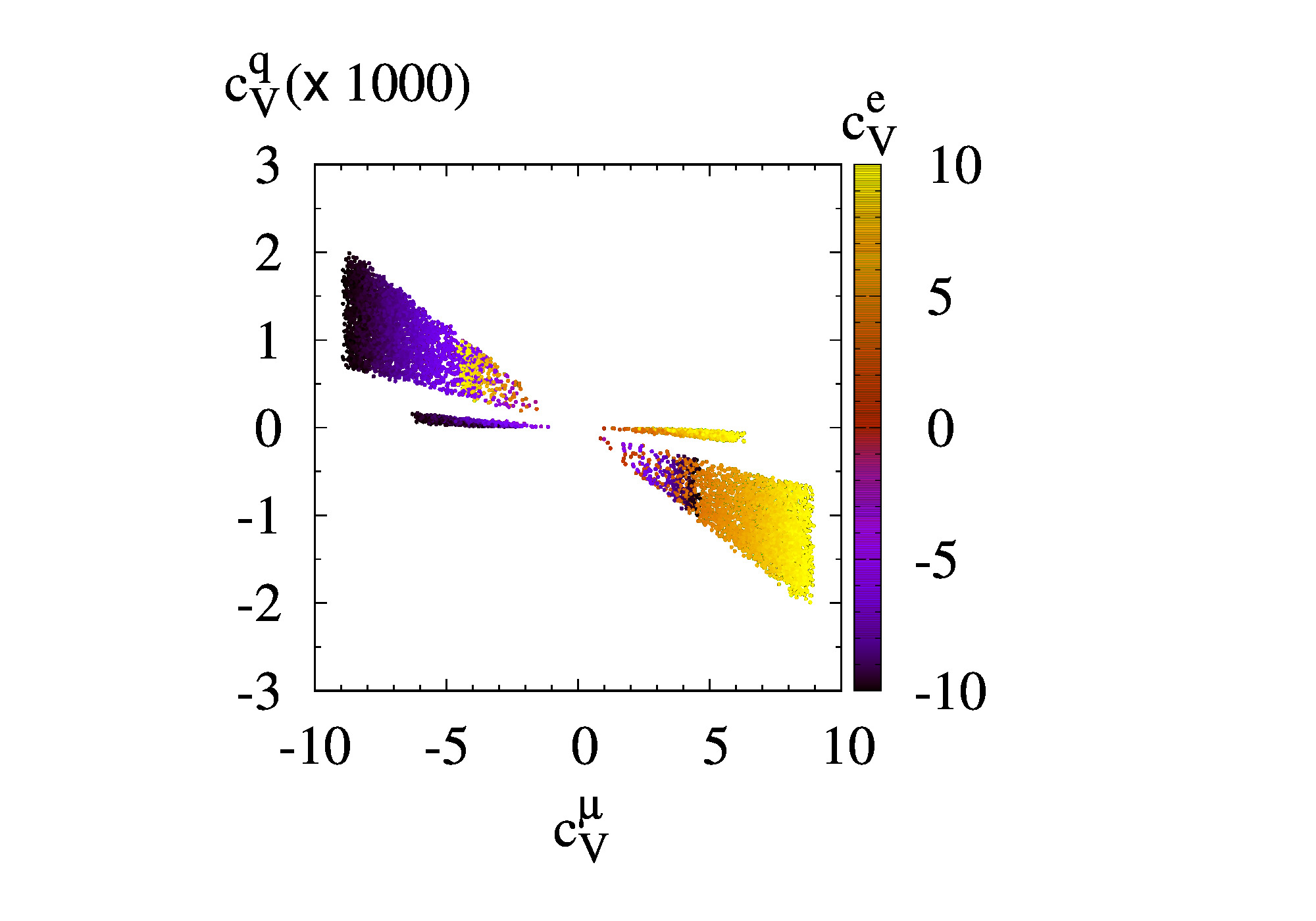}&
\hspace{-2cm}\includegraphics[scale=0.17]{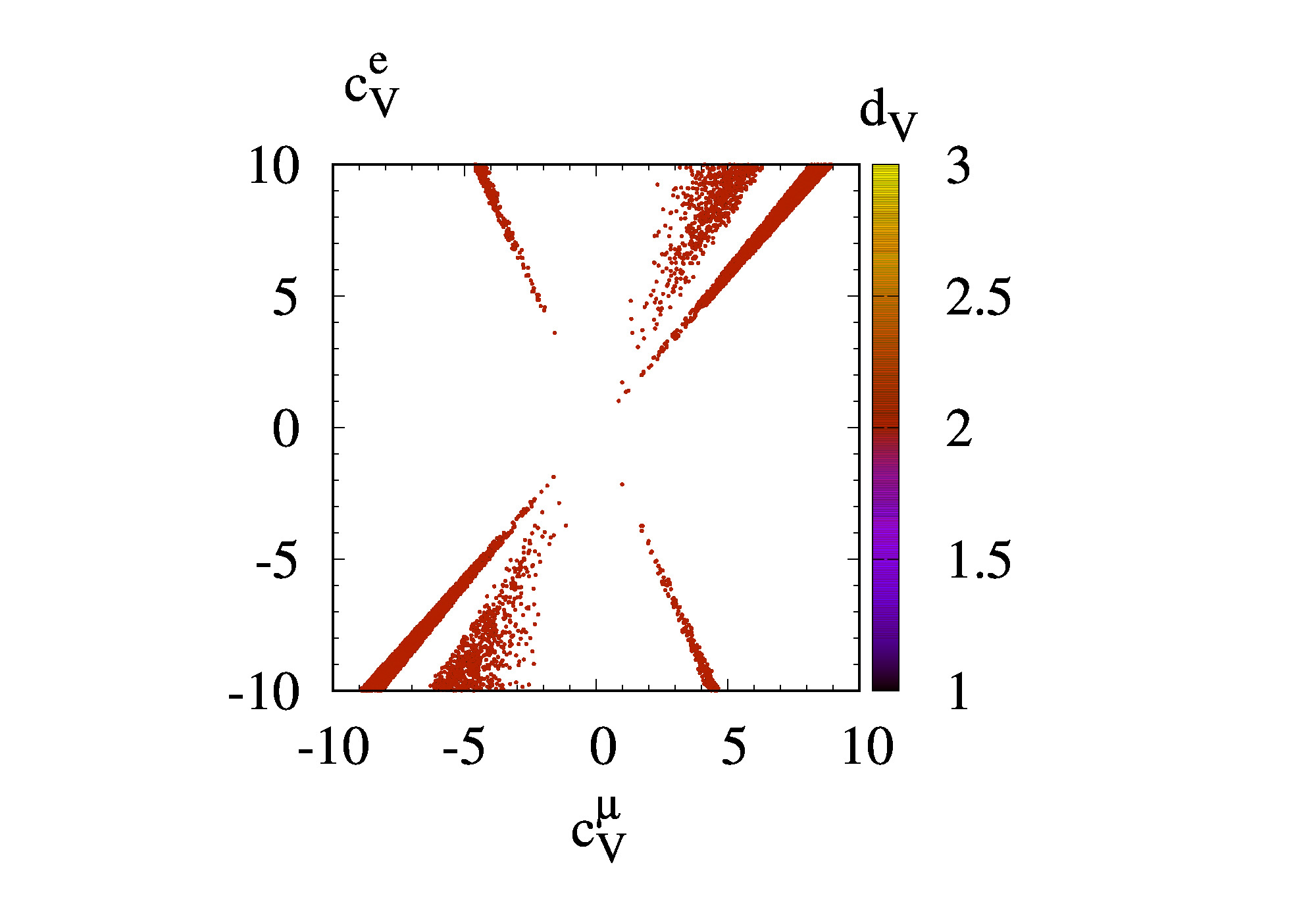}\\
\hspace{-2cm} (a) & \hspace{-2cm} (b) \\
\hspace{-2cm}\includegraphics[scale=0.17]{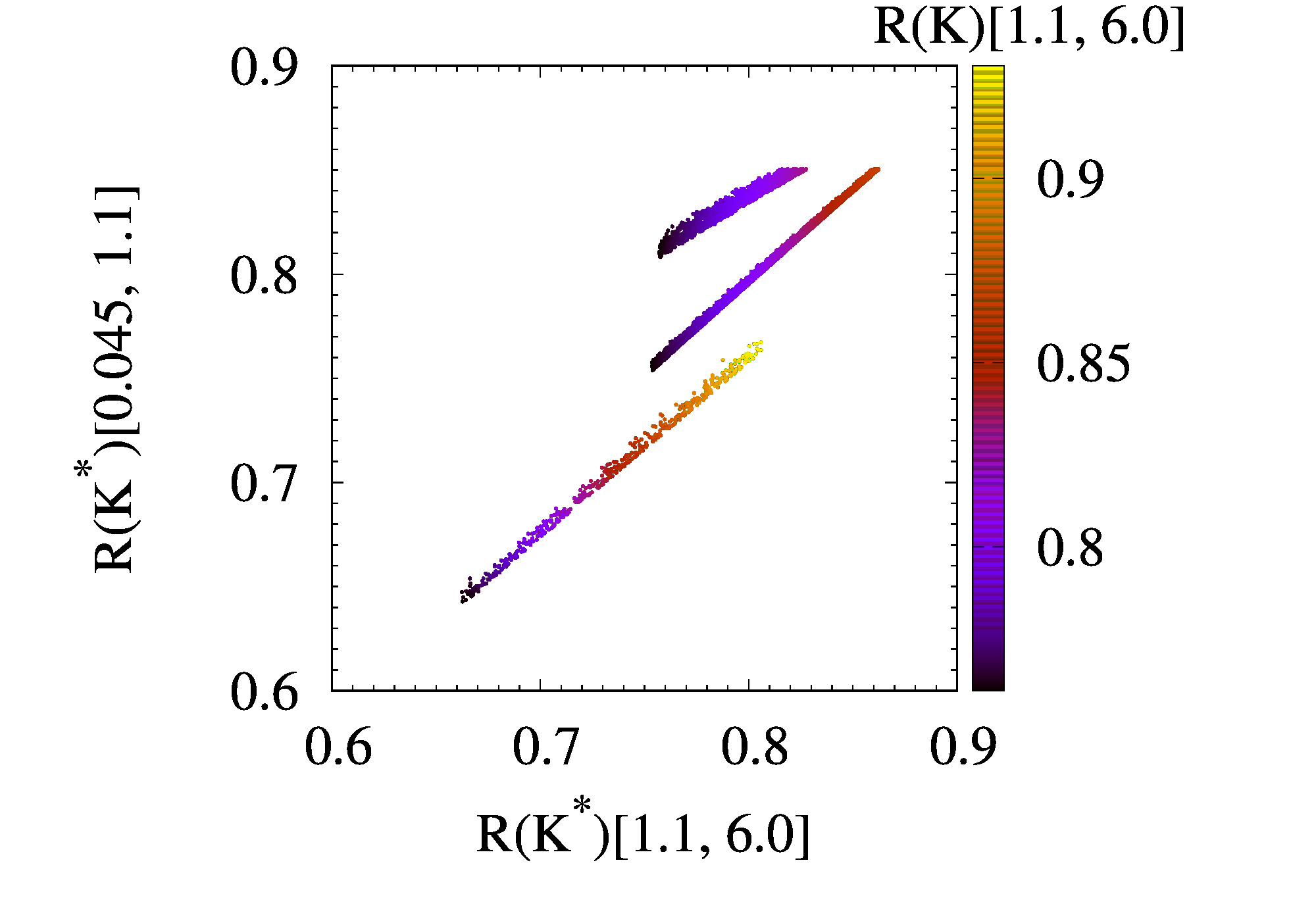}\\
\hspace{-2cm} (c) & 
\end{tabular}
\caption{\label{F_3d_cV} Allowed regions of (a) $c_V^q$ vs $c_V^\mu$ with respect to $c_V^e$,
(b) $c_V^e$ vs $c_V^\mu$ with respect to $d_V$, and
(c) $R(K^*)[0.045:1.1]$ vs $R(K^*)[1.1:6.0]$ with respect to $R(K)[1.1:6.0]$,
with all the constraints at the $2\sigma$ level.
}
\end{figure}

Figures \ref{F_3d_dV}-\ref{F_3d_cV} depict allowed regions for model parameters and observables
with all the constraints, $B_s$-$\Bbar_s$ mixing, $\Br(B_s\to\mu^+\mu^-)$, $\Delta a_{e,\mu}$, and $R(\Ks)$.
In Figs.\ \ref{F_3d_dV} (a) and (b) $c_V^\mu$ and $c_V^q$ behave in different ways with respect to $\LU$ as well as $d_V$.
It should be noted that only small vales of $|c_V^q|\lesssim 0.2\%$ are allowed.
The hint is given in Fig.\ \ref{F_dV} (b).
For $B_s\to\mu^+\mu^-$ decay there is an allowed region of small $d_V$ and small $c_V^q$,
which is favored by $\Delta a_\ell$ with very narrow $d_V\approx 2$.
Therefore the value of $|c_V^q|$ could be a good testing ground for the vector unparticle scenario. 
%
%
%
%\textcolor{red}{issue 1}
It should also be noted that the narrow range of $c_V^q$ is consistent with the previous analysis of
\cite{JPL1009}.
%\textcolor{red}{issue 1-END}
%
%
%
In both cases of $c_V^\mu$ and $c_V^q$, larger $\LU$ is allowed only for small values of $d_V$, as expected.
%
%
%     				End of Updates
%
%%%%%%%%%%%%%%%%%%%%%%%%%%%%%%%%%%%%%%%%%%%%%%%%%%%%%%%%%%%%%%
%%%%%%%%%%%%%%%%%%%%%%%%%%%%%%%%%%%%%%%%%%%%%%%%%%%%%%%%%%%%%%
%******* Woking Line *******
%%%%%%%%%%%%%%%%%%%%%%%%%%%%%%%%%%%%%%%%%%%%%%%%%%%%%%%%%%%%%%
%
\par
%----------------- Figure 5 ------------------------------------------------
%----------------------------------------------------------------------------
%
Figure \ref{F_3d_cV} shows allowed $c_V^{q,\ell}$ and $R(\Ks)$.
One can find that small values of $|c_V^q|$ are allowed (Fig.\ \ref{F_3d_cV} (a)) 
while small values of $|c_V^{e,\mu}|$ are disfavored (Fig.\ \ref{F_3d_cV} (b)), 
and the vector unparticles predict narrow band-like regions in $R(\Ks)$ (Fig.\ \ref{F_3d_cV} (c)).
These features could be tested in near future.
\par
A comment on $b\to s\gamma$ is in order. 
In Ref.\ \cite{He0805}, unparticle effects on $b\to s\gamma$ are studied and bounds on quark couplings are given.
Our couplings are larger than those of \cite{He0805} by $(v/m_q)$ where $v$ is the vacuum expectation value of the Higgs field.
In our language bounds on the product of the two couplings are roughly of order one at fixed $\LU=1~{\rm TeV}$,
which is consistent with our analysis.
One can expect that the bounds would go higher for larger values of $\LU$ because of the suppression factor $(v/\LU)^{d_V}$,
as discussed in \cite{JPL2012}.
\par
%
%
%
%%%%%%%%%%%%%%%%%%%%%%%%%%%%%%%%%%%%%%%%%%%%%%%%%%%%%%%
%\textcolor{red}{issue 4}
Our final comment is on the forward-backward asymmetry $A_{FB}$ of $B\to K^*\ell^+\ell^-$.
The SM predicts negative values of $A_{FB}$ for low $q^2$ and has zero crossing 
at $q^2=3.90\pm0.12$ GeV, and positive $A_{FB}$ for higher $q^2$. 
But experimental data show positive $A_{FB}$ for low $q^2$.
As discussed in \cite{Alok0912}, both $VA$ and $V'A'$ (chirality-flipped in the quark sector) interactions
are needed to fit the data.
In this analysis we only consider $VA$ interactions, so it would need to include new couplings involving 
$V'A'$ interactions to fit the $A_{FB}$.
%
%%%%%%%%%%%%%%%%%%%%%%%%%%%%%%%%%%%%%%%%%%%%%%%%%%%%%%%%%%%%%%%%%%%%%%
%%%%%%%%%%%%%%%%%%%%%%%%%%%%%%%%%%%%%%%%%%%%%%%%%%%%%%%%%%%%%%%%%%%%%%
\section{Conclusions}
%%%%%%%%%%%%%%%%%%%%%%%%%%%%%%%%%%%%%%%%%%%%%%%%%%%%%%%%%%%%%%%%%%%%%%
%
In this paper we showed that vector unparticles can successfully explain the $R(\Ks)$ anomalies and $a_\mu$
for a loosened condition of $1\le d_V\le 3$.
Compared to other NP models such as leptoquark or $Z'$, $\calU_V$ contributes to $R(\Ks)$ differently with the factor of
$s^{d_V-2}$ in the Wilson coefficients.
Constraints from the $B_s$-$\Bbar_s$ mixing are mild while those from $\Br(B_s\to\mu^+\mu^-)$ and $a_{e,\mu}$ restrict severely 
the model parameters.
Among the relevant couplings $c_V^q$ for quark sector can have only small values 
while other leptonic couplings are reluctant to have small absolute values.
If there would be some other constraints which require large values of $|c_V^q|\gg 0.2\%$ 
or small values of $|c_V^{e,\mu}|\ll 1$ 
then it could reject the validity of vector unparticles. 
We found that recently measured muon magnetic moment together with electron one puts stringent bounds on parameters and observables.
Our results are very promising in that only vector unparticles are considered and moderate values of model parameters 
could fit the experimental results. 
New interactions of chiality-flipped $V'A'$ could be included to explain other observables such as
forward-backward asymmetry of $B\to K^*\ell^+\ell^-$.
%Scalar unparticles could also be included to provide richer phenomenology and to explain future anomalies.

%%%%%%%%%%%%%%%%%%%%%%%%%%%%%%%%%%%%%%%%%%%%%%%%%%%%%%%%%%%%%%%%%%%%%%%%%%%%%%%%

%%%%%%%%%%%%%%%%%%%%%%%%%%%%%%%%%%%%%%%%%%%%%%%%%%%%%%%%%%%%%%%%%%%%%%

\end{document}